# Crystal chemistry aspects of the magnetically induced ferroelectricity in $TbMn_2O_5$ and $BiMn_2O_5$

L M Volkova and D V Marinin

Institute of Chemistry, Far Eastern Branch of the Russian Academy of Sciences
690022 Vladivostok, Russia

E-mail: volkova@ich.dvo.ru

**Abstract**
The origin of magnetic frustration was stated and the ions whose shift is accompanied by emerging magnetic ordering and ferroelectricity in $TbMn_2O_5$ and $BiMn_2O_5$ were determined on the basis of calculation of magnetic coupling parameters by using the structural data. The displacements accompanying the magnetic ordering are not polar, they just induce changes of bond valence (charge disordering) of Mn1 and Mn2, thus creating instability of the crystal structure. To approximate again the bond valence to the initial value (charge ordering) under magnetic ordering conditions is possible only due to polar displacement of Mn2 (or O1) and O4 ions along the $b$ axis that is the cause of ferroelectric transition.

## 1. Introduction

The problem of the crystal structure coupling with emerging magnetic ordering and electric polarization in multiferroics has been discussed rather intensively until recently. It was stated [1-12] that in multiferroics $RMn_2O_5$ (R - rare earth elements and Bi) the $Mn^{3+}$ and $Mn^{4+}$ ions are coupled by strong magnetic interactions competing with each other. Under applied magnetic field and at temperature about 40 K the antiferromagnetic ordering of Mn spins takes place, thus inducing the emerging of electric polarization along the $b$ axis. As was assumed in [6], polar atomic displacements resulted in the symmetry center disappearance and reduced the crystal symmetry from the $Pbam$ to the $Pb2_1m$ space group.

However, in spite of numerous attempts, direct experimental evidences of the presence of structural modifications accompanying the electric polarization in $RMn_2O_5$ have not been found yet. It must be related to the fact that up to present the structural studies of induced multiferroics $RMn_2O_5$ under high magnetic fields have not been conducted, since the required combination of high magnetic fields and X-ray diffraction equipment has only recently become available [13–17]. Besides, there exists another problem in studying the multiferroics structures. It is concerned with the fact, that in the process of studies of different samples of the same compound $RMn_2O_5$ by means of the X-ray powder diffraction methods one obtains a wide range of $Mn^{3+}$-O, $Mn^{4+}$-O and $R^{3+}$-O bond lengths. For example, in the paraelectric phase of three samples of $NdMn_2O_5$ [18-20] the difference of respective $Mn^{3+}$-O1, $Mn^{3+}$-O3, $Mn^{3+}$-O4, $Mn^{4+}$-O and Nd-O bond lengths attains 0.05 Å, 0.06 Å, 0.15 Å, 0.04-0.05 Å and 0.04-0.13 Å, respectively. Moreover, even the alternating along the $c$ axis long and short $Mn^{4+}$-$Mn^{4+}$ distances and the $Mn^{3+}$-$Mn^{3+}$ distance in the dimer vary over too wide range for heavy atoms: 2.93 - 2.99 Å, 2.71 - 2.77 Å and 2.86 - 2.90 Å, respectively. On the other hand, the difference in unit cell parameters is as small as 0.02 Å. Such structural differences may result from three factors: non-stoichiometry of the $NdMn_2O_5$ composition, low accuracy of the measurement method used or the structure instability (non-rigidity).

The emerging electric polarization at separation of the gravity centers of positive and negative charges might be the result of the displacement of cation itself from the polyhedron center or that of lighter oxygen anions or both types of ions. Reorientation of magnetic moments (antiferromagnetic (AF) – ferromagnetic (FM) transition) is also accompanied by displacements of intermediate ions in local space between magnetic ions. The ability of the surrounding cations' coordination to withstand such distortions would facilitate both reorientation of magnetic moments and emergence of dipole moments. We have analyzed the crystal structures of $Mn^{3+}$ and $Mn^{4+}$ from the Inorganic Crystal Structure Database (ICSD) (version 1.4.4, FIZ Karlsruhe,Germany, 2008-1), which were determined with the highest accuracy by means of x-ray single-crystal diffraction (the refinement converged to the residual factor (R) values R = 0.045-0.079) or neutron powder diffraction (R = 0.018-0.057) methods. The analysis has shown that the coordination polyhedra of both Jahn-Teller ion $Mn^{3+}$ and regular ion $Mn^{4+}$ are not "rigid" and meet the above requirement. The bond lengths and angles in the $Mn^{3+}$ and $Mn^{4+}$ polyhedra vary within a wide range in a random manner, since no regular increase of the bridge bonds lengths or decrease of the end bonds lengths was traced. For example, in the square pyramids $Mn^{3+}O_5$ coupled by common edges and vertices in the compounds $CaMn_2O_4$ [21], $KMnO_2$ [22], $Na_4Mn_2O_5$ [23] and $Ba_2Mn_2Si_2O_9$ [24] the lengths of the $Mn^{3+}$ bonds with oxygen atoms located in the pyramid vertex and base vary in the ranges 2.07-2.33 Å and 1.70-2.25 Å, respectively. The bond valence of $Mn^{3+}$ ions deviates significantly ($V_{Mn^{3+}}$ = 2.85-3.21) from the ideal value. In the octahedra $Mn^{4+}O_6$ coupled by common edges and vertices in the compounds $Pb_2MnO_4$ [25], $BaMn_3O_6$ [26], $Na_2Mn_3O_7$ [27] and $Ba_4Mn_3O_{10}$ [28], the $Mn^4$-O distances and valence bonds of $Mn^{4+}$ fall into the ranges 1.82-2.28 Å and 2.96-3.92, respectively. The ordered octahedral surrounding is observed only in high-symmetry crystals and has an enforced character.

The objective of the study is to determine which changes in the crystal structure could be the result of magnetic ordering of the frustrated antiferromagnetics $TbMn_2O_5$ and $BiMn_2O_5$ and why these changes can be the cause of emerging electrical polarization. To attain this objective, the sign and strength of magnetic interactions in the paraelectric phase of $TbMn_2O_5$ and $BiMn_2O_5$ will be calculated, and intermediate ions located in critical (or close to critical) positions of the local space between magnetic ions, deviations from which may cause reorientations of magnetic moments and emergence of magnetic ordering, will be found. The crystal structure of the magnetically ordered phase will be determined by varying these intermediate ions coordinates, and polar ions displacements resulting in ferroelectric transition at preserving magnetic ordering will be stated.

## 2. Method

The sign and strength of magnetic couplings in compounds were calculated by a new phenomenological method developed earlier [29] on the basis of structural data.

We have developed this method to estimate characteristics of magnetic interactions between magnetic ions located at any distances from each other. The main problem to be solved during the development of this method was to find a natural relation of the strength of the magnetic interactions and the type of magnetic moments ordering with crystal chemistry parameters in low-dimensional crystal compounds. For such a solution we used three widely spread conceptions on the nature of magnetic interactions discussed below. According to Kramers [30], the exchange couplings between magnetic ions separated by one or more diamagnetic groups are characterized by a significant contribution of non-magnetic ions electrons. The crystal chemistry aspect of the model of Goodenough-Kanamori-Anderson [31] unambiguously indicates the dependence of the interaction strength and the magnetic ions spins orientation type on the locations of intermediate ions. According to the polar model of Shubin-Vonsovsky [32], the determination of magnetic interactions characteristics should take into account not only the anions with valent bonds to magnetic ions, but also all the intermediate negatively or positively polarized atoms.

We have studied the relation of magnetic characteristics with crystal structure in low-dimensional compounds of d-elements on the experimental data provided in literature. As a result, we have found that the interaction between magnetic ions $M_i$ and $M_j$ emerges in the moment of crossing the boundary of the space between them by the intermediate ion $A_n$ ion. Here we take into account not only anions, which are valent-bound to the magnetic ions, but also all the intermediate negatively or positively ionized atoms, except cations of metals without unpaired electrons. The bound space region between the $M_i$ and $M_j$ ions along the bond line is defined as a cylinder whose radius is equal to that of these magnetic ions. If the magnetic ions are not identical, taking the radius of a smaller ion as the cylinder radius produced the best approximation to experimental results in all our cases under consideration. However, to make any final solution of this problem studies of a larger number of compounds are required. The strength of magnetic interactions and the ordering type of the magnetic moments in isolators are determined mainly by the geometrical arrangement and the size of the intermediate $A_n$ ion in the bound space region between two magnetic ions $M_i$ and $M_j$ (figure 1). The distance between magnetic ions, such as inside the low-dimensional fragment and between fragments, has an effect only on the contribution value, but does not determine the sign (type) of the contribution in the case of absence of a direct interaction contribution. The value of interaction into antiferromagnetic (AF) or ferromagnetic (FM) components of the interaction is maximal, if the intermediate ion is located in the central one-third of the space between the magnetic ions. To produce maximum contribution into the AF component the intermediate ion should be located near the axis, while for maximum contributing to the FM component, in contrast, it should be near the surface of a cylinder limiting the space area between magnetic fields.

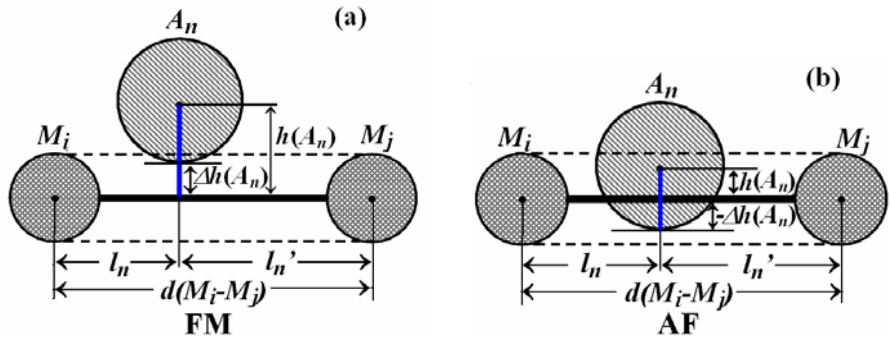

**Figure 1.** A schematic representation of the intermediate $A_n$ ion arrangement in the local space between magnetic ions $M_i$ and $M_j$ in cases when the $A_n$ ion initiates the emerging of the ferromagnetic (a) and antiferromagnetic (b) interactions. $\Delta h(A_n)$, $l_n$, $l_n'$ and $d(M_i\text{-}M_j)$ – the parameters determining the sign and strength of magnetic interactions.

If some intermediate ions enter the space between two magnetic ions, each of them, depending on the location, tends to orient the magnetic moments of these ions accordingly and makes a contribution to occurrence of AF or FM components of magnetic interaction. The sign and value of the strength of interaction $J_{ij}^S$ between the magnetic ions $M_i$ and $M_j$ are determined by the sum of these contributions $j_n^S$:

$$J_{ij}^S = \sum_n j_n^S . \qquad (1)$$

If $J_{ij}^S < 0$, the type of the magnetic moments ordering of $M_i$ and $M_j$ ions is antiferromagnetic, while if $J_{ij}^S > 0$, the type of the magnetic moments ordering is ferromagnetic.

The sign and strength of the $j_n^s$ contributions are determined by the degree of overlapping of the space between the magnetic ions by the intermediate $A_n$ ion, the degree of asymmetry of the $A_n$ ion location relatively to the bond line $M_i$-$M_j$ and the distance between magnetic fields $d(M_i$-$M_j)$:

$$j_n^s = \frac{\Delta h(A_n)\frac{l_n}{l_n'} + \Delta h(A_n)\frac{l_n'}{l_n}}{d(M_i - M_j)^2}, \quad (\text{if } l_n'/l_n < 2), \qquad (2)$$

and

$$j_n^S = \frac{\Delta h(A_n)\frac{l_n}{l_n'}}{d(M_i - M_j)^2}, \quad (\text{if } l_n'/l_n \geq 2). \qquad (3)$$

Here $\Delta h(A_n)$ is the difference between the distance $h(A_n)$ from the centre of $A_n$ ion up to the bond line $M_i$-$M_j$, while $r_{A_n}$ is the radius of the $A_n$ ion (figure 1):

$$\Delta h(A_n) = h(A_n) - r_{A_n}. \qquad (4)$$

This value characterizes the degree of space overlapping between the magnetic ions $M_i$ and $M_j$. If $\Delta h(A_n) < 0$, the $A_n$ ion overlaps (by $|\Delta h|$) the bond line $M_i$-$M_j$ and initiates the emerging contribution into the AF-component of magnetic interaction. If $\Delta h(A_n) > 0$, there remains a gap (the gap width $\Delta h$) between the bond line and the $A_n$ ion, and this ion initiates a contribution to the FM-component of magnetic interaction.

$l_n$ and $l_n'$ are the lengths of segments obtained by drawing a perpendicular from the center of the $A_n$ ion to the bond line $M_i$-$M_j$. Let us assume that $l_n \leq l_n'$; $l_n' = d(M_i - M_j) - l_n$. The $l_n'/l_n$ ratio characterizes the degree of asymmetry of the $A_n$ ion location relatively to the middle of the $M_i$-$M_j$ straight line. If $l_n'/l_n < 2.0$, the magnetic moments of both $M_i$ and $M_j$ ions will be under orientation effect of the intermediate $A_n$ ion and the $j_n^s$ calculation will have to be performed in accordance with formula (2). If $l_n'/l_n \geq 2$, the $A_n$ ion has an effect on orientation of the magnetic moment of the adjacent magnetic ion only, and the $j_n^s$ calculation will have to be performed in accordance with formula (3).

One should mention that during calculation of the $J_{ij}^s$ value it is necessary to additionally take into account the contribution from a direct interaction $j^D$, if the distance between the magnetic ions $d(M_i - M_j)$ is less than two diameters of these ions:

$$J_{ij}^S = \sum_n j_n^S + j^D. \qquad (5)$$

The analysis of the relation between magnetic and crystal chemistry parameters in low-dimensional copper compounds, in which $Cu^{2+}$ are located at short distances, has brought us to the above conclusion and allowed obtaining the expression for the $j^D$ calculation [29].

We based our considerations on the assumption that there exists some critical distance $D_c$ between the magnetic ions when the AF- and FM-contributions from a direct interaction are equal and eliminate each other. Deviation from $D_c$ results in AF-coupling in case of lower values and in FM-coupling in case of higher values. The value of the $j^D$ contribution is proportional to the deviation value ($d(M_i - M_j) - D_c$) and is in inverse proportion to the radii of magnetic ions $r_{M_n}$ and the distance between them $d(M_i - M_j)$:

$$j^D = \frac{d(M_i - M_j) - D_c}{r_{M_n} d(M_i - M_j)} \tag{6}$$

We have empirically found the $D_c$ value for $Cu^{2+}$ ions ($D_c$=2.88 Å).

To calculate the sign and value of the strength of magnetic interaction $J_{ij}^s$ we have developed the "MagInter" program. The program utilizes the expressions (1-4) obtained within the scopes of the method. The geometric parameters used in these expressions ($h(A_n)$, $\Delta h(A_n)$, $l_n$ and $l_n'$) are calculated from the interatomic distances and angles which, in turn, can be found through application of the program SELXTL [33]. The initial structural data format for the program (crystallographic parameters, atom coordinates) corresponds to that of the crystallographic information file (CIF) in the Inorganic Crystal Structure Database (ICSD) (FIZ Karlsruhe, Germany). Besides, in the calculations we used the ionic radii determined by Shannon [34].

The determined parameters of magnetic interactions are displayed only in cases when there are no restrictions for their simultaneous existence due to geometric configurations in the magnetic ions sublattices. The presence of specific configurations of magnetic ions results in geometric frustration of magnetic interactions. For non-stoichiometric compounds one should additionally take into account the presence of vacancies.

*2.1. Critical positions of intermediate ions*

There exist several critical positions of intermediate $A_n$ ions when even a slight deviation from them could result in reorientation of magnetic moments (AF – FM transition) and/or dramatic change of the magnetic interaction strength. It appears important to note that under effects of temperature, pressure, magnetic field and others the ions in crystal structure could undergo displacement. That is why during prediction of possible changes in the sign and strength of magnetic interactions one should take into account not only the ions located exactly at critical positions, but also those in adjacent areas.

The following intermediate ions positions can be considered as critical:

(a) $h(A_n) = r_M + r_{A_n}$: the distance $h(A_n)$ from the $A_n$ ion center to the bond line $M_i$-$M_j$ is equal to the sum of the $M$ and $A_n$ ions radii. The $A_n$ ion reaches the surface of a cylinder of the radius $r_M$, limiting the space area between the magnetic ions $M_i$ and $M_j$. In this case the $A_n$ ion does not induce the emerging of a magnetic interaction. However, at slight decrease of $h(A_n)$ (the $A_n$ ion displacement inside this area) there emerges a strong FM-interaction between magnetic ions.

(b) $h(A_n) = r_{A_n}$ ($\Delta h(A_n) = 0$): the distance $h(A_n)$ from the center of the $A_n$ ion to the bond line $M_i$-$M_j$ is equal to the $A_n$ ion radius (the $A_n$ reaches the bond line $M_i$-$M_j$). In this case the interaction between magnetic fields disappears. However, at slight decrease of $h(A_n)$ (overlapping of the bond line by the $A_n$ ion) there emerges a weak AF-interaction, while at slight increase of $h(A_n)$ (formation of a gap between the $A_n$ ion and the bond line $M_i$-$M_j$) there emerges a weak FM-interaction.

(c) $l_n'/l_n = 2$: the $A_n$ ion is located at the boundaries of the central one-third of the space between magnetic fields. In this case the insignificant displacement of the $A_n$ ion to the center in parallel to the bond line $M_i$-$M_j$ results in a dramatic increase of the magnetic interaction strength.

In case when there are several intermediate $A_n$ ions between the magnetic ions $M_i$ and $M_j$, the following critical positions are possible:

(d) When the ratio between the sums of $j_n^s$ contributions into the AF- and FM-components of the interaction becomes close to 1, the interaction between the magnetic ions $M_i$ and $M_j$ is weak, and a slight displacement of even one of the intermediate $A_n$ ions could result in its complete disappearance or the AF-FM transition.

(e) When even one of the intermediate $A_n$ ions is in a critical position of (a) or (c) type, the contribution to AF- or FM-components of the interaction could undergo dramatic changes because of even slight displacement of these ions and, therefore, cause changes of respective scale in the interaction strength and reorientation of magnetic ions spins.

The sign and strength of magnetic couplings in paraelectric phases of TbMn$_2$O$_5$ and BiMn$_2$O$_5$ as well as in magnetically ordered nonpolar and polar models of these compounds were calculated with using the program "MagInter". The structural data for TbMn$_2$O$_5$ at room temperature [35] and BiMn$_2$O$_5$ at room temperature [36] and T = 100K [11] were taken for calculations and models development. Besides, in the above calculations we used the ionic radii determined by Shannon [34] ($r_{Mn^{3+}}$ = 0.53 Å (coordination number CN is equal to 5), $r_{Mn^{4+}}$ = 0.58 Å (CN = 6), $r_{O^{2-}}$ = 1.40 Å). The contribution from the direct interaction between manganese ions $j^D$ was not taken into account, since all the distances between manganese in these compounds are longer than two Mn ions diameters. The bond valences of manganese ions $V_{Mn^{3+}}$ and $V_{Mn^{4+}}$ were calculated by Brese and O'Keeffe [37]. The structural parameters, bond distances and bond valences of Mn$^{3+}$ and Mn$^{4+}$ ions in TbMn$_2$O$_5$, BiMn$_2$O$_5$ and models are presented in Tables 1 and 2. The parameters of the main magnetic interactions in these compounds and models are presented in Tables 3 and 4.

## 3. Results and Discussion

*3.1. Characterization of magnetic interactions and their competition in paraelectric phases of TbMn$_2$O$_5$ and BiMn$_2$O$_5$*

The compounds TbMn$_2$O$_5$ [35] and BiMn$_2$O$_5$ [11, 36] in the paraelectric phase crystallize in a centrosymmetrical orthorhombic space group *Pbam* and have two types of magnetic ions: Mn1 (Mn$^{4+}$, S=3/2) in distorted oxygen octahedra Mn$^{4+}$O$_6$ and Mn2 (Mn$^{3+}$, S=2) in distorted square pyramids Mn$^{3+}$O$_5$ (figures 2(a) and (b)). The Mn$^{4+}$O$_6$ octahedra are coupled alternately by common edges O2-O2 and O3-O3 into a linear chain along the *c* axis. The Mn$^{3+}$O$_5$ pyramids are coupled into dimers by the common edge O1-O1. These dimers couple the octahedra chains along the *a* axis through the ions O3 located in the pyramids vertices and along the *b* axis through the ions O4 located in the pyramids bases.

According to our calculations, in the paraelectric phases of TbMn$_2$O$_5$ and BiMn$_2$O$_5$ at room temperature and T = 100K the respective magnetic couplings between manganese ions are of the same sign and differ insignificantly in strength (tables 3 and 4; figures 2(d), 3(a) and (b)). The compound BiMn$_2$O$_5$ at room temperature and that of 100 K are referred to as Bi-RT and Bi-LT, respectively.

Strong AF *J*1 and *J*2 couplings alternate in the linear chain along the *c* axis (figure 2(d)). The *J*1 coupling is weaker than the *J*2 coupling (the ratio of intrachain couplings *J*2/*J*1 = 1.45 (1.27 and 1.31) in Tb(Bi-RT and Bi-LT)-systems). The main contribution into the AF-components of the *J*1 and *J*2 couplings is provided by two O2 ions (the value of contribution from two O2 ions: $2j_{O2}$ = -0.059 (-0.058 and -0.058) Å$^{-1}$) and two O3 ions ($2j_{O3}$ = -0.077 (-0.070 and -0.072) Å$^{-1}$), respectively. Higher values of these AF-contributions result from the fact that the O2 and O3 ions are located in the central one-third part of the space between magnetic ions ($l'/l$) = 1. However, the O2 and O3 ions positions are close to the critical position "b" (see Section 2), since the distances from these ions to the bond line Mn1-Mn1 ($h$(O2) = 1.27 Å and $h$(O3) = 1.26-1.27 Å) are slightly less than the oxygen ion radius ($r_{O^{2-}}$ = 1.40 Å), i.e. the ions O2 and O3 slightly overlap the bond line Mn1-Mn1.

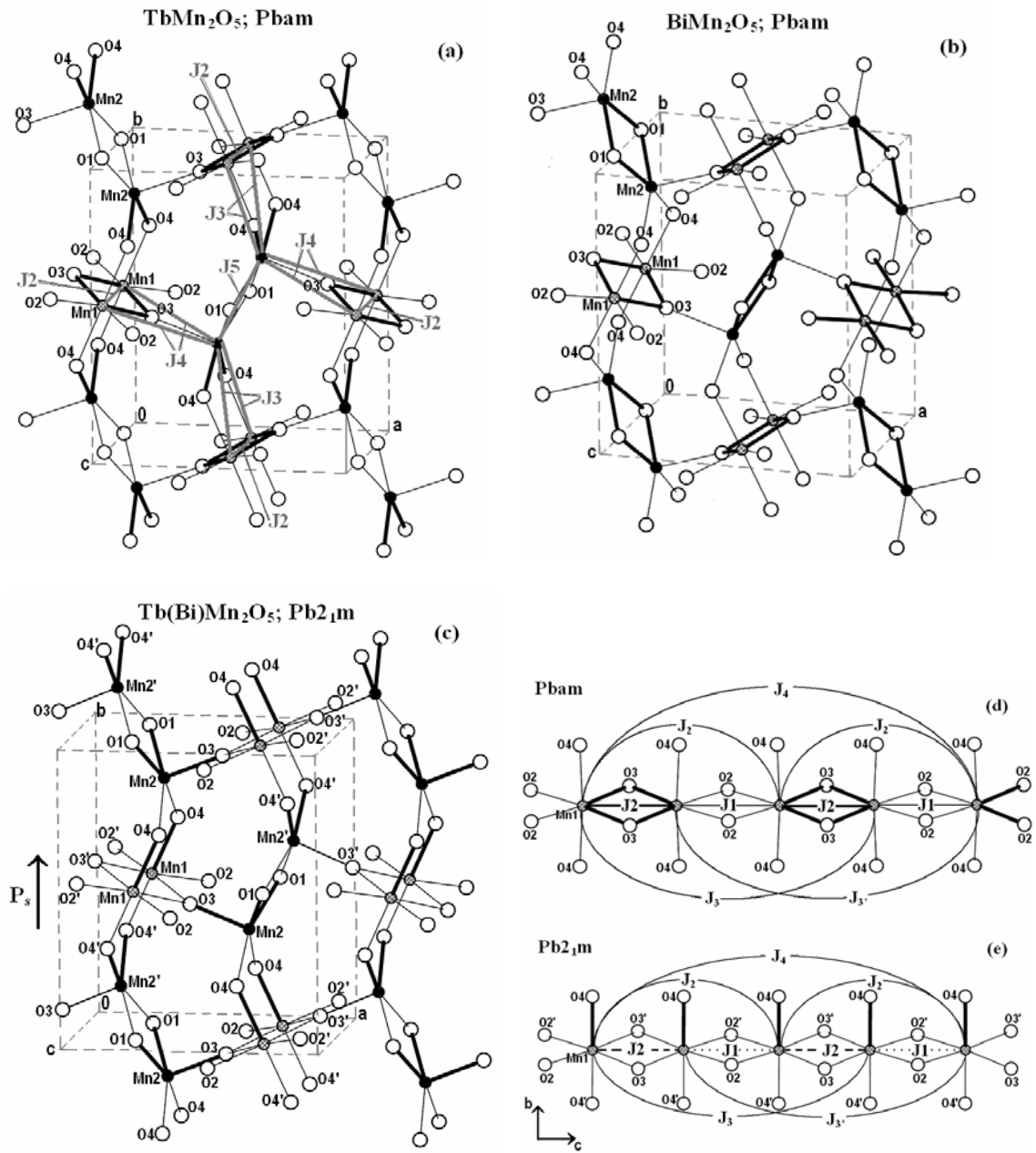

**Figure 2**. The relation of Mn-O bond lengths in Mn1 and Mn2 polyhedra: the crystal structure of the paraelectric phase of TbMn$_2$O$_5$ (a) and BiMn$_2$O$_5$ (b) and magnetically ordered ferroelectric model Tb(Bi)Mn$_2$O$_5$ at polar displacements of Mn2, Mn2', O4 and O4' (c). The linear chain along the $c$ axis and the $J_n$ coupling in the paraelectric phase (d) and for the magnetically ordered ferroelectric model (e). Thick and thin lines refer to short and long Mn-O bonds, respectively.

Slight O2 and O3 ions displacement perpendicular to the bond line Mn1-Mn1 could control the spin orientation of the $J_1$ and $J_2$ couplings. The contributions of O2 and O3 ions into the $J_1$ and $J_2$ couplings disappear, if $h$(O2) and $h$(O3) increase by 0.13-0.14 Å and reach 1.40 Å. During further removal of these ions from the bond line the character of their contribution changes into ferromagnetic one and, as a result, the $J$1 and $J$2 couplings undergo the AF→FM transition. Aside from two O2 ions, the $J$1 coupling space contains four O4 ions which initiate emerging of slight contributions into the FM-component of this interaction, since they are removed from the middle of the Mn1-Mn1 straight line to the Mn1 ions. Moreover, the O4 ions are located near the boundary of the interaction space (the critical position 'a') and, in case of their displacement perpendicular to the chain (along the *b*-axis) by as little as 0.01-0.02 Å from the line Mn1-Mn1, they leave the boundaries and cannot participate in the $J$1 coupling formation. However, the O4 ions have a leading role in couplings with the second ($J_2$), third ($J_3$) and fourth ($J_4$) neighbors in the linear chain along the *c*-axis, since they appear in the central one-third part of these interactions space.

**Table 1.** Structural parameters, bond distances and bond valences of Mn1 and Mn2 ions ($V_{Mn1}$, $V_{Mn2}$) in TbMn$_2$O$_5$

| TbMn$_2$O$_5$ | Alonso [35] | M-DO[a] Model | P-Mn2[b] Model | | P-O1[c] Model | | Wang [7, 8][d] | |
|---|---|---|---|---|---|---|---|---|
| Space group | Pbam | Pbam | Pb2$_1$m | | Pb2$_1$m | | Pb2$_1$m **N26** | |
| a (Å) | 7.3251 | 7.317 | 7.317 | | 7.317 | | 7.3014 | |
| b (Å) | 8.5168 | 8.506 | 8.506 | | 8.506 | | 8.5393 | |
| c (Å) | 5.6750 | 5.660 | 5.660 | | 5.660 | | 5.6056 | |
| Mn$^{4+}$O$_6$ | Mn1 | Mn1 | Mn1 | | Mn1 | | Mn1 | |
| Mn1-O2(x2) (Å) | 1.954 | 2.013 | 2.013 | | 2.013 | | 1.923(O2, O2') | |
| Mn1-O3(x2) (Å) | 1.847 | 2.013 | 2.013 | | 2.013 | | 1.871(O3, O3') | |
| Mn1-O4(x2) (Å) | 1.912 | 1.908 | 1.791(O4); 1.808 (O4') | | 1.791(O4); 1.826 (O4') | | 1.923(O4); 1.937(O4') | |
| $V_{Mn1}$ | 4.01 | 3.30 | 3.75 | | 3.70 | | 3.96 | |
| Mn$^{3+}$O$_5$ | Mn2 | Mn2 | Mn2 | Mn2' | Mn2 | Mn2' | Mn2 | Mn2' |
| Mn2-O1(x2) (Å) | 1.927 | 1.928 | 1.915(O1) | 1.942(O1) | 1.915(O1) | 1.942(O1) | 1.919(O1) | 1.928(O1) |
| Mn2-O3(x1) (Å) | 2.021 | 1.805 | 1.797(O3) | 1.813(O3') | 1.805(O3) | 1.805(O3') | 1.908(O3) | 1.912(O3') |
| Mn2-O4(x2) (Å) | 1.890 | 1.855 | 1.956(O4) | 1.915(O4') | 1.942(O4) | 1.915(O4') | 1.914(O4) | 1.911(O4') |
| $V_{Mn2}$ | 3.18 | 3.70 | 3.40 | 3.40 | 3.42 | 3.42 | 3.29 | 3.26 |
| Mn1: x | 0 | 0 | 0.2500 | | 0.2500 | | 0.2501 | |
| y | ½ | ½ | 0.5000 | | 0.5000 | | 0.5003 | |
| z | 0.2618 | 0.2557 | 0.2557 | | 0.2557 | | 0.2558 | |
| Mn2(Mn2'): x | 0.4120 | 0.4114 | 0.6614(0.1614) | | 0.6614(0.1614) | | 0.6512 (0.1516) | |
| y | 0.3510 | 0.3505 | 0.3529(0.1519) | | 0.3505(0.1495) | | 0.3558 (0.1456) | |
| z | ½ | ½ | ½ (½) | | ½ (½) | | ½ (½) | |
| O1: x | 0 | 0 | 0.2500 | | 0.2500 | | 0.2508 | |
| y | 0 | 0 | 0.0000 | | -0.0024 | | 0.0002 | |
| z | 0.2710 | 0.2710 | 0.2710 | | 0.2710 | | 0.2709 | |
| O2(O2'): x | 0.1617 | 0.1808 | 0.4308(0.9308) | | 0.4308(0.9308) | | 0.4146 (0.9148) | |
| y | 0.4463 | 0.4463 | 0.4463(0.0537) | | 0.4463(0.0537) | | 0.4480 (0.0517) | |
| z | 0 | 0 | 0(0) | | 0(0) | | 0 (0) | |
| O3(O3'): x | 0.1528 | 0.1838 | 0.4338(0.9338) | | 0.4338(0.9338) | | 0.4060 (0.9071) | |
| y | 0.4324 | 0.4324 | 0.4324(0.0676) | | 0.4324(0.0676) | | 0.4329 (0.0655) | |
| z | ½ | ½ | ½ (½) | | ½ (½) | | ½ (½) | |
| O4(O4'): x | 0.3973 | 0.3973 | 0.6473(0.1473) | | 0.6473(0.1473) | | 0.6477 (0.1459) | |
| y | 0.2062 | 0.2062 | 0.1911(0.3067) | | 0.1911(0.3043) | | 0.2077 (0.2919) | |
| z | 0.2483 | 0.2550 | 0.2550(0.7450) | | 0.2550(0.7450) | | 0.2438 (0.7579) | |

[a] Magnetic ordering and elimination of dipole moments of Mn1O$_6$ octahedra.
[b] Spontaneous polarization along the *b* axis accompanied by the Mn2, Mn2', O4 and O4' ions displacement.
[c] Spontaneous polarization along the *b* axis accompanied by the O1, O4 and O4' ions displacement.
[d] Coordinates of all TbMn$_2$O$_5$ atoms from [7,8] are displaced by ¼ along the *x* axis to preserve the structural motif.

The $J_2$ couplings are ferromagnetic and weaker than the AF $J$1 couplings (the ratio $J_2/J1$ is equal to -0.47(-0.44 and -0.44) in Tb(Bi-RT and Bi-LT)-systems), since the O4 ions reduce comparatively large AF-contribution $j_{Mn1}$ ($j_{Mn1}$ = -0.033(-0.032 and -0.032) Å$^{-1}$ in Tb(Bi-RT and Bi-LT)-systems) of the intermediate Mn1 ion and small AF-contributions of two intermediate ions O2 and O3. However, in case of the O4 ions leaving the interaction space the $J_2$ couplings will undergo the phase transition FM→AF ($J_2/J1 = 0.66$) and compete with the $J1$ and $J2$ couplings.

**Table 2.** Structural parameters, bond distances and bond valences of Mn1 and Mn2 ions ($V_{Mn1}$, $V_{Mn2}$) in BiMn$_2$O$_5$

| BiMn$_2$O$_5$ | | Munoz [36] | Granado [11] | M-DO[a] Model | P-Mn2[b] Model | | P-O1[c] Model | |
|---|---|---|---|---|---|---|---|---|
| Space group | | Pbam | Pbam | Pbam | Pb2$_1$m | | Pb2$_1$m | |
| $a$ (Å) | | 7.5608 | 7.54116 | 7.54116 | 7.54116 | | 7.54116 | |
| $b$ (Å) | | 8.5330 | 8.52994 | 8.52994 | 8.52994 | | 8.52994 | |
| $c$ (Å) | | 5.7607 | 5.75437 | 5.75437 | 5.75437 | | 5.75437 | |
| Mn$^{4+}$O$_6$ | | Mn1 | Mn1 | Mn1 | Mn1 | | Mn1 | |
| Mn1-O2(x2) (Å) | | 1.968 | 1.961 | 2.047 | 2.047 | | 2.047 | |
| Mn1-O3(x2) (Å) | | 1.870 | 1.872 | 2.047 | 2.047 | | 2.047 | |
| Mn1-O4(x2) (Å) | | 1.910 | 1.922 | 1.922 | 1.765(O4); 1.818 (O4') | | 1.765(O4); 1.864 (O4') | |
| $V_{Mn1}$ | | 3.88 | 3.86 | 3.07 | 3.61 | | 3.52 | |
| Mn$^{3+}$O$_5$ | | Mn2 | Mn2 | Mn2 | Mn2 | Mn2' | Mn2 | Mn2' |
| Mn2-O1(x2) (Å) | | 1.899 | 1.897 | 1.897 | 1.894(O1) | 1.961(O1) | 1.894(O1) | 1.961(O1) |
| Mn2-O3(x1) (Å) | | 2.085 | 2.086 | 1.820 | 1.803(O3) | 1.838(O3') | 1.820(O3) | 1.820(O3') |
| Mn2-O4(x2) (Å) | | 1.929 | 1.916 | 1.916 | 2.075(O4) | 1.952(O4') | 2.038(O4) | 1.952(O4') |
| $V_{Mn2}$ | | 3.06 | 3,11 | 3.54 | 3.14 | 3.16 | 3.19 | 3.20 |
| Mn1: | x | ½ | ½ | ½ | 0.7500 | | 0.7500 | |
| | y | 0 | 0 | 0 | 0.0000 | | 0.0000 | |
| | z | 0.2613 | 0.2596 | 0.2596 | 0.2596 | | 0.2596 | |
| Mn2(Mn2'): | x | 0.4074 | 0.40755 | 0.40755 | 0.65755(0.15755) | | 0.65755(0.15755) | |
| | y | 0.3516 | 0.35091 | 0.35091 | 0.35691(0.15509) | | 0.35091(0.14909) | |
| | z | ½ | ½ | ½ | ½ (½) | | ½ (½) | |
| O1: | x | 0 | 0 | 0 | 0.2500 | | 0.2500 | |
| | y | 0 | 0 | 0 | 0.0000 | | -0.0060 | |
| | z | 0.2866 | 0.2876 | 0.2876 | 0.2795 | | 0.2795 | |
| O2(O2'): | x | 0.1553 | 0.1567 | 0.1750 | 0.4250(0.9250) | | 0.4250(0.9250) | |
| | y | 0.4440 | 0.4453 | 0.4453 | 0.4453(0.0547) | | 0.4453(0.0547) | |
| | z | 0 | 0 | 0 | 0(0) | | 0(0) | |
| O3(O3'): | x | 0.1440 | 0.1437 | 0.1809 | 0.4309(0.9309) | | 0.4309(0.9309) | |
| | y | 0.4241 | 0.4243 | 0.4243 | 0.4243(0.0757) | | 0.4243(0.0757) | |
| | z | ½ | ½ | ½ | ½ (½) | | ½ (½) | |
| O4(O4'): | x | 0.3856 | 0.3866 | 0.3866 | 0.6366(0.1366) | | 0.6366(0.1366) | |
| | y | 0.1995 | 0.2018 | 0.2018 | 0.1810(0.3120) | | 0.1810(0.3059) | |
| | z | 0.2539 | 0.2525 | 0.2525 | 0.2525(0.7455) | | 0.2525(0.7455) | |

[a] Magnetic ordering and elimination of dipole moments of Mn1O$_6$ octahedra.
[b] Spontaneous polarization along the $b$ axis accompanied by the Mn2, Mn2', O4 and O4' ions displacement.
[c] Spontaneous polarization along the $b$ axis accompanied by the O1, O4 and O4' ions displacement.

The $J_3$ and $J_{3'}$ couplings between the third neighbors in the chain are not equivalent (tables 3 and 4; figure 2(d)). The $J_3$ coupling is very weak ($J_3/J1 = 0.04$), belongs to the AF-type and does not compete with the nearest couplings of the chain. Alternatively, the $J_{3'}$ coupling is of the FM-type ($J_{3'}/J1 = -0.47$ (-45 and -45) in Tb(Bi-RT and Bi-LT)-systems) and competes with the nearest $J1$ and $J2$ couplings. However, in the local space of the $J_3$ and $J_{3'}$ couplings the locations of O4 ions providing the highest contribution to the FM-component of these interactions is critical in regard to displacements in two directions: perpendicular (critical position "a") and parallel (critical position "c") to the chain. Removal of the O4 ions beyond the interaction space boundary at slight displacement (by 0.02 Å) along the $b$-axis increases the strength of the AF $J_3$ coupling ($J_3/J1 = 0.27(0.28)$), transforms the $J_{3'}$ coupling from the FM-type into the AF-type ($J_{3'}/J1 = 0.75(0.72$ and $0.072$) in Tb(Bi-RT and Bi-LT)-systems), and, thus, eliminates its competition with the nearest-neighbor interactions. Significant changes in the $J_3$ and $J_{3'}$ couplings at slight displacements of the O4 ions in parallel to the chain result from the fact that these ions are located near ($l'/l$ approximately equal 2) the boundary of the central one-third part of local space of these interactions. In the $J_3$ coupling the O4 ions are located beyond, while in the $J_{3'}$ coupling – inside the central part of the interaction space. The displacement (by 0.02(0.04) Å in Tb(Bi)-systems) of the O4 ions into the central part of the $J_3$ coupling and the accompanying removal of these ions from the central part of the $J_{3'}$ coupling result in the transition of the $J_3$ coupling into the FM-state ($J_3/J1 = -1.13(-1.02$ and $-1.04)$ in Tb(Bi-RT and Bi-LT)-systems) and the $J_{3'}$ coupling into the AF-state ($J_{3'}/J1 = 0.57(0.51)$ in Tb(Bi)-systems). In this case there is still competition in the chain, but exclusively due to the $J_3$ couplings.

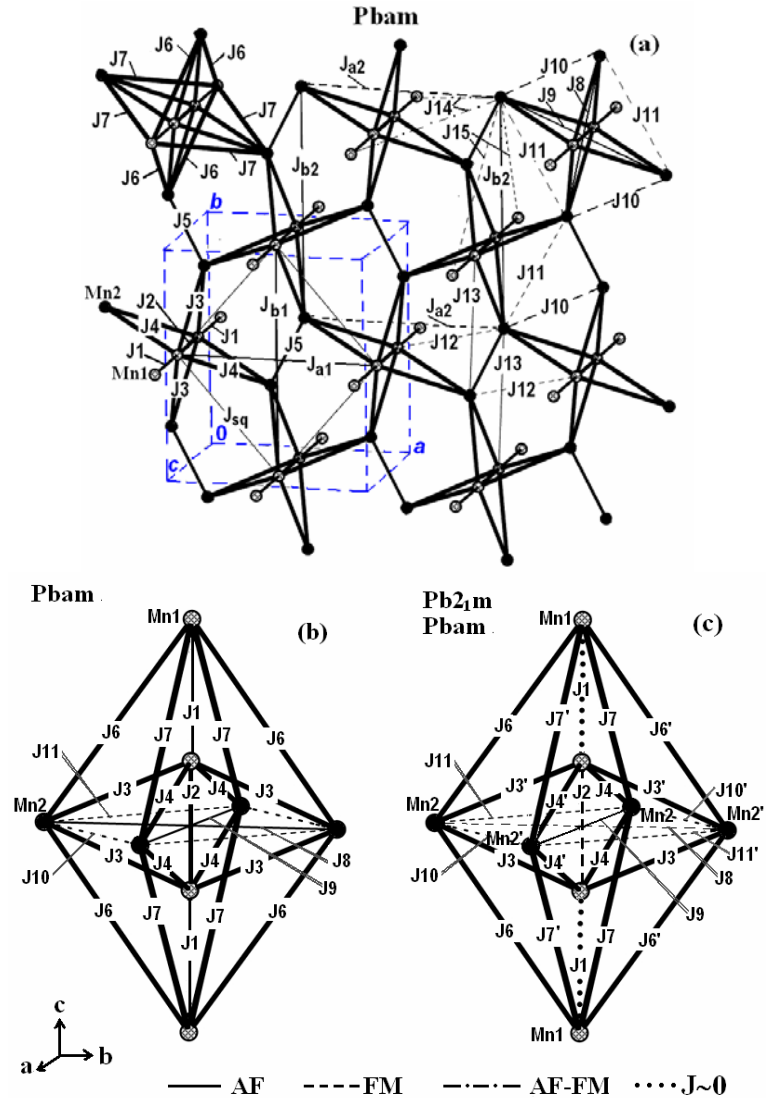

**Figure 3**. (a) The sublattice of $Mn^{3+}$(Mn2) and $Mn^{4+}$(Mn1) and the coupling $J_n$ in frustrated paraelectric phase of $Tb(Bi)Mn_2O_5$. Fragment of sublattice: change of the coupling $J_n$ parameters at transition from frustrated paraelectric state (b) to magnetically ordered ferroelectric state (Models M-DO, P-Mn2 and P-O1) (c). The thickness of lines shows the strength of the $J_n$ coupling. AF, FM couplings and absence of coupling are indicated by solid, dashed and dotted lines, respectively. The possible FM→AF transitions are shown by stroke in dashed lines.

**Table 3.** Parameters of magnetic couplings in TbMn$_2$O$_5$ calculated on the basis of structural data.

| TbMn$_2$O$_5$ | Alonso [35] | M-DO[a] Model | P-Mn2[b] Model | P-O1[c] Model | Wang [7, 8] |
|---|---|---|---|---|---|
| Space group | Pbam | Pbam | Pb2$_1$m | Pb2$_1$m | Pb2$_1$m |
| a (Å) | 7.3251 | 7.317 | 7.317 | 7.317 | 7.3014 |
| b (Å) | 8.5168 | 8.506 | 8.506 | 8.506 | 8.5393 |
| c (Å) | 5.6750 | 5.660 | 5.660 | 5.660 | 5.6056 |
| d(Mn1-Mn1) (Å) | 2.971 | 2.895 | 2.895 | 2.895 | 2.868 |
| $J_1$ (Å$^{-1}$) | -0.059[d], -0.053 | 0 | 0 | 0 | -0.058[d], -0.055 |
| d(Mn1-Mn1) (Å) | 2.704 | 2.765 | 2.765 | 2.765 | 2.738 |
| $J_2$ (Å$^{-1}$) | -0.077 | 0.033 | 0.033 | 0.033 | -0.066 |
| d(Mn1-Mn1) (Å) | 5.675 | 5.660 | 5.660 | 5.660 | 5.606 |
| $J_2 = J_{c1}$ (Å$^{-1}$) | -0.039[d], 0.025 | 0.032 | 0.018 | 0.019 | -0.039[d], -0.006 |
| d(Mn1-Mn1) (Å) | 8.379 | 8.425 | 8.425 | 8.425 | 8.343 |
| $J_3$ (Å$^{-1}$) | -0.016[de], -0.002[e] -0.016[df], 0.060[f] | 0.007 | 0.004 | 0.005 | -0.016[d], 0.022 |
| d(Mn1-Mn1) (Å) | 8.646 | 8.555 | 8.555 | 8.555 | 8.473 |
| $J_{3'}$ (Å$^{-1}$) | -0044[de], -0.030[e] -0044[df], 0.025[f] | 0.036 | 0.022 | 0.023 | -0.045[d], -0.037 |
| d(Mn1-Mn1) (Å) | 11.350 | 11.320 | 11.320 | 11.320 | 11.211 |
| $J_4$ (Å$^{-1}$) | -0.021[d], 0.000 | 0.012 | 0.008 | 0.008 | -0.021[d], -0.010 |
| d(Mn1-Mn2) (Å) | 3.344 | 3.350 | 3.368 (3.332) | 3.350 (3.350) | 3.407 (3.401) |
| $J3$ ($J3'$) (Å$^{-1}$) | -0.071 | -0.078 | -0.081 (-0.082) | -0.082 (-0.081) | -0.071 (-0.068) |
| d(Mn1-Mn2) (d(Mn1-Mn2')) (Å) | 3.542 | 3.548 | 3.541 (3.556) | 3.548 (3.548) | 3.460 (3.466) |
| $J4$ ($J4'$) (Å$^{-1}$) | -0.085 | -0.095 | -0.095 (-0.094) | -0.095 (-0.094) | -0.093 (-0.093) |
| d(Mn2-Mn2) (Å) | 2.847 | 2.855 | 2.855 | 2.855 | 2.863 |
| $J5$ (Å$^{-1}$) | -0.050 | -0.051 | -0.051 | -0.051 | -0.057 |
| d(Mn1-Mn2) (d(Mn1-Mn2')) (Å) | 5.295 | 5.254 | 5.266 (5.242) | 5.254 (5.254) | 5.262 (5.257) |
| $J6$ ($J6'$) (Å$^{-1}$) | -0.081 | -0.081 | -0.071 (-0.075) | -0.073 (-0.075) | -0.084 (-0.084) |
| d(Mn1-Mn2) (d(Mn1-Mn2')) (Å) | 5.423 | 5.383 | 5.378 (5.388) | 5.383 (5.383) | 5.296 (5.300) |
| $J7$ ($J7'$) (Å$^{-1}$) | -0.094 | -0.087 | -0.086 (-0.086) | -0.086 (-0.086) | -0.095 (-0.095) |
| d(Mn2-Mn2) (Å) | 6.116 | 6.102 | 6.102 | 6.102 | 6.233 |
| $J8$ (Å$^{-1}$) | -0.020[d], -0.018 | 0.005 | 0.005 | 0.005 | -0.015 |
| d(Mn1-Mn1) (Å) | 5.617 | 5.610 | 5.610 | 5.610 | 5.617; 5.619 |
| $J_{sq}$ (Å$^{-1}$) | -0.007 | -0.008 | -0.009 | -0.009 | -0.007; -0.007 |
| d(Mn1-Mn1) (Å) | 7.325 | 7.317 | 7.317 | 7.317 | 7.301 |
| $J_{a1}$ (Å$^{-1}$) | -0.047 | -0.046 | -0.046 | -0.046 | -0.046 |
| d(Mn1-Mn1) (Å) | 8.517 | 8.506 | 8.506 | 8.506 | 8.539 |
| $J_{b1}$ (Å$^{-1}$) | -0.040 | -0.040 | -0.040 | -0.040 | -0.039 |
| d(Mn2-Mn2) (d(Mn2'-Mn2')) (Å) | 7.325 | 7.317 | 7.317 (7.317) | 7.317 (7.317) | 7.301 (7.301) |
| $J_{a2}$ ($J_{a2'}$) (Å$^{-1}$) | 0.023 | 0.021 | 0.018 (0.019) | 0.019 (0.018) | 0.023 (0.023) |
| d(Mn2-Mn2) (d(Mn2'-Mn2')) (Å) | 8.517 | 8.506 | 8.506 (8.506) | 8.506 (8.506) | 8.539 (8.539) |
| $J_{b2}$ ($J_{b2'}$) (Å$^{-1}$) | -0.024 | 0.012 | 0.012 (0.012) | 0.012 (0.012) | -0.025 (-0.025) |
| d(Mn2-Mn2) (d(Mn2'-Mn2')) (Å) | 5.675 | 5.660 | 5.660 (5.660) | 5.660 (5.660) | 5.606 (5.606) |
| $J_{c2}$ ($J_{c2'}$) (Å$^{-1}$) | -0.003 | 0.027 | 0.029 (0.029) | 0.029 (0.029) | 0.029 (0.030) |

[a] Magnetic ordering and elimination of dipole moments of Mn1O$_6$ octahedra.
[b] Spontaneous polarization along the *b* axis accompanied by the Mn2, Mn2', O4 and O4' ions displacement.
[c] Spontaneous polarization along the *b* axis accompanied by the O1, O4 and O4' ions displacement.
[d] During calculation of the $J$n coupling the contribution from an intermediate ion located in the critical position 'a' was not taken into account.
[e] During calculation of the $J$n coupling the formula (3) is taken, since some intermediate ions are localized in the critical position 'c'.
[f] During calculation of the $J$n coupling the formula (2) is taken, since some intermediate ions are localized in the critical position 'c'.

The $J_4$ coupling strength attains zero ($J_4/J1 = 0$(0 and -0.02) in Tb(Bi-RT and Bi-LT)-systems), since the sum of contributions of three intermediate Mn1 ions, four O2 ions and two O3 ions into the AF-component of the interaction is approximately equal to the sum of eight O4 ions into the interaction FM-component. In Tb and Bi-HT systems these contributions eliminate each other while in Bi-RT-system the FM-contribution is slightly higher than the AF-contribution. However, in case of slight displacement of the O4 ions from the line -Mn1-Mn1- beyond the boundary of the interaction space the $J_4$ coupling will be transformed into comparatively strong AF coupling ($J_4/J1 = 0.36$-$0.40$) and compete with the nearest couplings in the chain.

**Table 4.** Parameters of magnetic couplings in BiMn$_2$O$_5$ calculated on the basis of structural data.

| BiMn$_2$O$_5$ | Munoz[36] | Granado[11] | M-DO[a] Model, | P-Mn2[b] Model | P-O1[c] Model |
|---|---|---|---|---|---|
| Space group | Pbam | Pbam | Pbam | Pb2$_1$m | Pb2$_1$m |
| a (Å) | 7.5608 | 7.54116 | 7.54116 | 7.54116 | 7.54116 |
| b (Å) | 8.5330 | 8.52994 | 8.52994 | 8.52994 | 8.52994 |
| c (Å) | 5.7607 | 5.75437 | 5.75437 | 5.75437 | 5.75437 |
| d(Mn1-Mn1) (Å) | 3.011 | 2.988 | 2.988 | 2.988 | 2.988 |
| $J1$ (Å$^{-1}$) | -0.058[d], -0.055 | -0.058[d], -0.055 | 0.003 | 0.002 | 0.002 |
| d(Mn1-Mn1) (Å) | 2.750 | 2.767 | 2.767 | 2.767 | 2.767 |
| $J2$ (Å$^{-1}$) | -0.070 | -0.072 | 0.057 | 0.057 | 0.057 |
| d(Mn1-Mn1) (Å) | 5.761 | 5.754 | 5.754 | 5.754 | 5.754 |
| $J_2 = J_{c1}$ (Å$^{-1}$) | -0.038[d], 0.024 | -0.038[d], 0.026 | 0.033 | 0.017 | 0.020 |
| d(Mn1-Mn1) (Å) | 8.511 | 8.521 | 8.521 | 8.521 | 8.521 |
| $J_3$ (Å$^{-1}$) | -0.016[d,e], -0.002[e] -0.016[d,f], 0.056[f] | -0.016[d,e], -0.002[e] -0.016[d,f], 0.057[f] | 0.008 | 0.005 | 0.005 |
| d(Mn1-Mn1) (Å) | 8.771 | 8.742 | 8.742 | 8.742 | 8.742 |
| $J_{3'}$ (Å$^{-1}$) | -0.042[de], -0.028[e] -0.042[df], 0.024[f] | -0.043[de], -0.028[e] -0.043[df], 0.025[f] | 0.040 | 0.023 | 0.025 |
| d(Mn1-Mn1) (Å) | 11.521 | 11.509 | 11.509 | 11.509 | 11.509 |
| $J_4$ (Å$^{-1}$) | -0.020[d], 0.000 | -0.021[d], 0.001 | 0.014 | 0.009 | 0.010 |
| d(Mn1-Mn2) (Å) | 3.374 | 3.370 | 3.370 | 3.416 (3.325) | 3.370 (3.370) |
| $J3$ ($J3'$) (Å$^{-1}$) | -0.067 | -0.067 | -0.064 | -0.066 (-0.070) | -0.068 (-0.067) |
| d(Mn1-Mn2) (d(Mn1-Mn2')) (Å) | 3.603 | 3.602 | 3.602 | 3.585 (3.621) | 3.602 (3.602) |
| $J4$ ($J4'$) (Å$^{-1}$) | -0.080[d], -0.076 | -0.080[d], -0.076 | -0.091 | -0.095 (-0.089) | -0.090 (-0.090) |
| d(Mn2-Mn2) (Å) | 2.894 | 2.901 | 2.901 | 2.901 | 2.901 |
| $J5$ (Å$^{-1}$) | -0.082 | -0.085 | -0.085 | -0.062 | -0.062 |
| d(Mn1-Mn2) (d(Mn1-Mn2')) (Å) | 5.360 | 5.343 | 5.343 | 5.372 (5.315) | 5.343 (5.343) |
| $J6$ ($J6'$) (Å$^{-1}$) | -0.071 | -0.073 | -0.071 | -0.058 (-0.067) | -0.060 (-0.067) |
| d(Mn1-Mn2) (d(Mn1-Mn2')) (Å) | 5.507 | 5.493 | 5.493 | 5.481 (5.505) | 5.493 (5.493) |
| $J7$ ($J7'$) (Å$^{-1}$) | -0.093 | -0.093 | -0.085 | -0.084 (-0.084) | -0.084 (-0.084) |
| d(Mn2-Mn2) (Å) | 6.162 | 6.147 | 6.147 | 6.147 | 6.147 |
| $J8$ (Å$^{-1}$) | -0.019 | -0.019 | 0.011 | 0.011 | 0.011 |
| d(Mn1-Mn1) (Å) | 5.700 | 5.693 | 5.693 | 5.693 | 5.693 |
| $J_{sq}$ (Å$^{-1}$) | -0.008 | -0.008 | -0.009 | -0.010 | -0.009 |
| d(Mn1-Mn1) (Å) | 7.561 | 7.541 | 7.541 | 7.541 | 7.541 |
| $J_{a1}$ (Å$^{-1}$) | -0.041 | -0.040 | -0.040 | -0.042 | -0.042 |
| d(Mn1-Mn1) (Å) | 8.533 | 8.530 | 8.530 | 8.530 | 8.530 |
| $J_{b1}$ (Å$^{-1}$) | -0.037 | -0.036 | -0.038 | -0.039 | -0.039 |
| d(Mn2-Mn2) (d(Mn2'-Mn2')) (Å) | 7.561 | 7.541 | 7.541 | 7.541 (7.541) | 7.541 (7.541) |
| $J_{a2}$ ($J_{a2'}$) (Å$^{-1}$) | 0.021 | 0.022 | 0.024 | 0.019 (0.022) | 0.021 (0.021) |
| d(Mn2-Mn2) (d(Mn2'-Mn2')) (Å) | 8.533 | 8.530 | 8.530 | 8.530 (8.530) | 8.530 (8.530) |
| $J_{b2}$ ($J_{b2'}$) (Å$^{-1}$) | -0.027 | -0.027 | -0.021 | -0.020 (-0.021) | -0.020 (-0.020) |
| d(Mn2-Mn2) (d(Mn2'-Mn2')) (Å) | 5.761 | 5.754 | 5.754 | 5.754 (5.754) | 5.754 (5.754) |
| $J_{c2}$ ($J_{c2'}$) (Å$^{-1}$) | -0.001 | -0.002 | 0.031 | 0.033 (0.034) | 0.033 (0.033) |

[a] Magnetic ordering and elimination of dipole moments of Mn1O$_6$ octahedra.
[b] Spontaneous polarization along the b axis accompanied by the Mn2, Mn2', O4 and O4' ions displacement.
[c] Spontaneous polarization along the b axis accompanied by the O1, O4 and O4' ions displacement.
[d] During calculation of the $J$n coupling the contribution from an intermediate ion located in the critical position 'a' was not taken into account.
[e] During calculation of the $J$n coupling the formula (3) is taken, since some intermediate ions are localized in the critical position 'c'.
[f] During calculation of the $J$n coupling the formula (2) is taken, since some intermediate ions are localized in the critical position 'c'.

The AF intra-dimer $J5$ coupling is formed under effect of two intermediate O1 ions localized in the center between the Mn2 ions ($l'/l=1$) at a distance $h(O1) = 1.300(1.229$ and $1.222)$ Å from the O1 ion center to the straight line Mn2-Mn2. Each O1 ion contributes -0.025(-0.041 and -0.042) Å$^{-1}$ to emerging of the AF-component of the $J5$ coupling in Tb(Bi-RT and Bi-LT)-system. As a result, the $J5$ coupling in Tb-system is markedly weaker ($J5^{Tb}/J5^{Bi-HT} = 0.61$, $J5^{Tb}/J5^{Bi-LT} = 0.59$) than in Bi-system. The O1 ions are located in the critical position 'b' and could control the strength and sign of the $J5$ coupling. The intra-dimer coupling could disappear, if $h$(O1) increases up to 1.40 Å, or change the sign to opposite, if $h$(O1) surpasses this value.

In the $ab$ plane (figures 2(a), 3(a) and (b)) the linear chains –Mn1-Mn1- and dimers Mn2-Mn2 are coupled by the strong AF $J3$ (along the $b$ axis) and $J4$ (along the $a$ axis) couplings ($J1 = 0.7(0.8)J2 = 1.1(0.7)J5 = 0.7(0.8)J3 = 0.6(0.7)J4$ in Tb(Bi)-system). The $J3$ coupling emerges mainly under effect of one intermediate O4 ion localized in the central one-third part of the space ($l'/l \approx 1$) between the ions Mn1 and Mn2 at a distance $h$(O4) = 0.905 - 0.919 Å from the straight line Mn1-Mn2 and makes a substantial contribution $j_{O4}$ ($j_{O4}$ = -0.089(-0.085) Å$^{-1}$ in Tb(Bi)-system) to the interaction AF-component. The $J3$ coupling value slightly reduces due to small contribution of two O3 ions and one O4 ion into this interaction ferromagnetic component. The $J4$ coupling is formed under effect of the O3 ion ($l_{O3}'/l_{O3}$ = 1.11(1.14); $h$(O3) = 0.777(0.819) Å in Tb(Bi)-system) making a substantial AF-contribution ($j_{O3}$ = -0.100(-0.091) Å$^{-1}$) and an insignificant FM-contribution from the O1, O2, O3 and O4 ions.

Beside the above interactions, more remote strong AF $J6$ (along the $b$ axis) and $J7$ (along the $a$ axis) couplings take place between linear chains and dimers. The $J7$ coupling is a dominating interaction ($J7/J1$ = 1.77(1.69) in Tb(Bi)-system) in all the compounds under consideration. The contribution (from -0.101 Å$^{-1}$ to -0.102 Å$^{-1}$) into the AF-component of this interaction emerges under effect of one O2 ion located in the central one-third part of the interaction space ($l_{O2}'/l_{O2}$ = 1.78-1.71), almost on the straight line Mn1-Mn2 ($h$(O2) = 0.099-0,121 Å). The $J6$ coupling is weaker than the $J7$ coupling ($J6/J7$ 0.86(0.76 and 0.78) in Tb(Bi-RT and Bi-LT)-systems)), since the O4 ion, which initiates its formation, is located farther from the line Mn1-Mn2 ($h$(O4)= 0.391-0.493 Å; $l_{O4}'/l_{O4}$ = 1.86-1.87).

One should emphasize that, unlike the intra-chain $J1$, $J2$ and $J_n$ couplings and intra-dimer $J5$ coupling, the $J3$, $J4$, $J6$ and $J7$ couplings are stable, since they do not contain intermediate ions in critical locations. Reorientation of the $J3$ and $J6$ couplings spins requires a substantial distortion of the crystal structure (O4 displacement along the $x$ axis >0.5 Å) while reorientation of the $J4$ and $J7$ couplings spins is impossible within the frames of given crystal structure.

All strong AF $J1$, $J2$, $J3$, $J4$, $J6$ and $J7$ couplings form geometrically frustrated isosceles $J2J3J3$ and $J2J4J4$ triangles or distorted $J1J3J6$ and $J1J4J7$ triangles (figures 3(a) and (b)). Besides, the $J3$, $J4$, $J6$ and $J7$ couplings compete with weaker inter-dimer AF $J8$ and $J9$ couplings in the triangles $J8J3J3$, $J8J6J6$, $J9J4J4$ and $J9J7J7$.

The Mn1 ions sublattice can be presented as an antiferromagnetic square lattice in the $ab$ plane with competing interactions along the side ($J_{sq}$) and diagonal of the square ($J_{a1}$ and $J_{b1}$). The value of frustration ratio of the second-neighbor (diagonal) coupling to the nearest-neighbor (side) coupling is significantly higher than the critical value ($\alpha$=1/2) and falls into the range 4.5-6.7.

The arrangement of AF Mn2-Mn2 magnetic dimers in the $ab$ planes is similar to that of Cu-Cu dimers in the compound SrCu$_2$(BO$_3$)$_2$ [38], however, unlike SrCu$_2$(BO$_3$)$_2$, the inter-dimer couplings ($J10$ and $J11$) in Tb(Bi)Mn$_2$O$_5$ are ferromagnetic. There is a weak competition between one AF $J5$ coupling and two weak FM $J10$ and $J11$ couplings, which form a distorted triangle $J5J10J11$, since these couplings are not equal in strength ($J5$ = -2.4(-5.1 and -5.3)$J10$ = -7.1(-9.1 and -9.4)$J11$ in Tb(Bi-RT and Bi-LT)-systems). Besides, the FM $J10$ and $J11$ couplings compete with the AF inter-dimer couplings $J8$ and $J9$ in the distorted triangles $J8J10J11$ and $J9J10J11$.

Comparatively strong AF $J_{a1}$ and $J_{b1}$ couplings exist between the Mn1 ions, which are located through the elementary unit parameters along the $a$ and $b$ axes, while along the $c$ axis the $J_{c1}$ ($J_{c1} \equiv J_2$) coupling is ferromagnetic, but could undergo the phase transition FM→AF, as shown above. Similar couplings between the Mn2 ions are significantly weaker do not always have the same sign: the $J_{b2}$ and $J_{c2}$ couplings are antiferromagnetic, and the $J_{c2}$ coupling could undergo the phase transition AF→FM (intermediate O1 and O4 ions are located in the critical position 'b'), while the $J_{a2}$ coupling is ferromagnetic.

One should mention that the parameters of magnetic couplings, which we calculated in accordance with the structural data of the Wang polar model [7, 8] (table 1), are similar to those of the frustrated paraelectric phase TbMn$_2$O$_5$ (table 3).

Thus, the paraelectric phases of TbMn$_2$O$_5$ and BiMn$_2$O$_5$ are frustrated antiferromagnetics. In linear chains along the *c*-axis, the nearest AF $J$1 and $J$2 couplings compete with the next-to-nearest-neighbor ($J_n$) couplings. In addition, the $J$1 and $J$2 couplings compete with all strong AF $J$3, $J$4, $J$6 and $J$7 couplings between linear chains and dimers. However, the AF intra-chain $J$1 and $J$2 couplings are unstable and could be eliminated or transformed into the FM-state, resulting in magnetic ordering, even at slight displacement of intermediate O2, O3 and O4 ions from the bond line -Mn1-Mn1-. Further we will consider the relation of magnetic ordering with emerging of electric polarization.

*3.2. Necessary conditions for emerging ferroelectricity in TbMn$_2$O$_5$ and BiMn$_2$O$_5$*

The crystal structure of Tb(Bi)Mn$_2$O$_5$ is not typical for ferroelectrics, since in the paraelectric phase it contains atomic groups in the forms of Mn1$^{4+}$O6 octahedra and Mn2$^{3+}$O5 pyramids having constant dipole moments; however, coupling in these groups is of substantially ionic character. The dipoles in these groups are oriented in different directions. The gravity centers of positive and negative centers in Mn$^{4+}$O$_6$ distorted octahedra and Mn$^{3+}$O$_5$ pyramids do not coincide (tables 1 and 2; figures 2(a) and (b)). In Mn1O$_6$ octahedra the Mn1 is displaced (by ~0.07(~0.06 Å) in Tb(Bi)-systems) along the *c* axis from the octahedron center to the O3-O3 edge that makes the Mn1-O3 distances in the octahedron equatorial plane shorter (by ~0.1Å) than the Mn1-O2 distances. The distances to the O4 ions in octahedron axial vertices are approximately equal to average value of long Mn1-O2 and short Mn-O3 bond lengths. The dipole moments of octahedra in the chain are oriented antiparallel and directed along the *c* axis.

In dimers the Mn2 are displaced from the Mn$^{3+}$O$_5$ pyramids centers to the O4-O4 edges in TbMn$_2$O$_5$ and in opposite direction to the common O1-O1 edge in BiMn$_2$O$_5$. As a result, the Mn2-O4 bond lengths in TbMn$_2$O$_5$ are shorter (by 0.04 Å) than the Mn2-O1 bond lengths while in BiMn$_2$O$_5$, in opposite, they are longer (by 0.03(0.02) in Bi-RT(Bi-LT)-systems). The dipole moments of pyramids in dimers are oriented antiparallel and directed along the *b* axis, however, the direction of dipoles in TbMn$_2$O$_5$ is changed by 180° relatively to that in BiMn$_2$O$_5$.

For emerging spontaneous polarization in RMn$_2$O$_5$ it is necessary to perform ordering of dipoles of the Mn$^{4+}$O$_6$ octahedra and Mn$^{3+}$O$_5$ pyramids in the same direction or eliminate the dipole moment in one of the groups while ordering dipoles in another group. Changing of orientation of dipole moments of octahedra and pyramids is possible due to displacement of specific ions, because couplings in these dipole groups are not "rigid". It is not the rigidity of coordination polyhedra of Mn$^{3+}$ and Mn$^{4+}$ (see section 1) that allows changing the bond lengths in wide ranges, however, to attain the compound stable state it is necessary to approximate the sum of bond valences to the cation ideal valence. The presence of common O3 and O4 ions for chains and dimers cause interrelation of any structural changes in these fragments.

The process of emerging electric polarization will be considered for two non-centrosymmetrical space groups *Pba*2 and *Pb*2$_1$*m*, which comprise subgroups of the space group *Pbam* and belong to the *mm*2 symmetry class. The atoms positions for Tb(Bi)Mn$_2$O$_5$ in the space group *Pbam* and its subgroups *Pba*2 and *Pb*2$_1$*m* are presented in table 5.

At a first glance, the space group *Pba*2 seems to be the most suitable for ferroelectric transition, since it has the same reflection conditions with the centrosymmetrical space group *Pbam* and is indistinguishable from it on X-ray pictures. However, in this group the *c*-axis serves as the polar axis while the magnetoelectric measurements show that spontaneous polarization occurs along the *b*-direction. Nevertheless, it was of interest for us to find out why the *b*-direction had become more preferable than the *c*-direction. In the Pba2 space group the positions of Mn1(Mn$^{4+}$), O1 and O4 ions are split into two kinds of sites.

**Table 5.** Comparison of the positions in the space group *Pbam* and its subgroups *Pba*2 and *Pb*2$_1$*m*.

| *Pbam* ($D^9_{2h}$) N55 | | *Pba*2 ($C^8_{2v}$) N32 | | *Pb*2$_1$*m* ($C^2_{2v}$) N26 | |
|---|---|---|---|---|---|
| 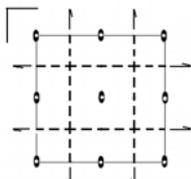 | | 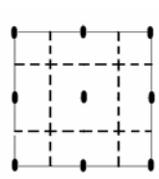 | | 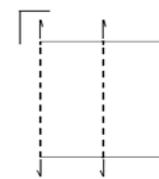 | |
| Reflection conditions 0kl: k=2n; h0l: h=2n; h00: h=2n; 0k0: k=2n | | Reflection conditions 0kl: k=2n; h0l: h=2n; h00: h=2n; 0k0: k=2n | | Reflection conditions 0kl: k=2n; 0k0: k=2n | |
| Position Atom | Coordinates | Position Atom | Coordinates | Position Atom | Coordinates |
| 8*i* O4 | x, y, z; -x, -y, z; -x+½, y+½, z; x+½, -y+½, z; -x, -y, -z; x, y, -z; x+½, -y+½, -z; -x+½, y+½, -z | 4*c*: O4 (x, y, z) O4' (-x, -y, -z) | x, y, z; -x, -y, z; -x+½, y+½, z; x+½, -y+½, z | 4*c*: O4 (x+1/4, y, z); O4' (-x+1/4, -y, -z) | x, y, z -x, y+1/2, z x, y, -z -x, y+1/2, -z |
| 4*h* Mn2, O3 | x, y, ½; -x, -y, ½; -x+½, y+½, ½; x+½, -y+½, ½ | 4*c*: Mn2, O3 | x, y, ~½; -x, -y, ~½; -x+½, y+½, ~½; x+½, -y+½, ~½ | 2*b*: Mn2, O3 (x+1/4, y, ½); Mn2', O3' (-x+1/4, -y, ½ ) | x, y, ½ -x, y+½, ½ |
| 4*g* R, O2 | x, y, 0; -x, -y, 0; -x+½, y+½, 0; x+½, -y+½, 0 | 4*c*: R, O2 | x, y, ~0; -x, -y, ~0; -x+½, y+½, ~0; x+½, -y+½, ~0 | 2*a*: R, O2 (x+1/4, y, 0) R', O2' (-x+1/4, -y, 0) | x, y, 0; -x, y+1/2, 0 |
| 4*f* Mn1 | 0, ½, z; ½, 0, z; 0, ½, -z; ½, 0, -z | 2*b*: Mn1 (z) Mn1' (-z) | 0, ½, z; ½, 0, z | 4*c*: Mn1 (~1/4, ~½, z) | x, y, z -x, y+1/2, z x, y, -z -x, y+1/2, -z |
| 4*e* O1 | 0, 0, z; ½, ½, z; 0, 0, -z; ½, ½, -z | 2*a*: O1 (z) O1' (-z) | 0, 0, z; ½, ½, z | 4*c*: O1(~1/4, ~0, z) | x, y, z -x, y+1/2, z x, y, -z -x, y+1/2, –z |

The *b*-axis serves as the polar axis in the non-centrosymmetrical space group *Pb*2$_1$*m*. Transition into this space group is concerned with emerging of additional reflections, however, they were not found during the crystals studies by X-ray and neutron diffraction methods in absence of external field. In the space group *Pb*2$_1$*m* the positions of Mn2, O2, O3 and O4 ions are split into two kinds of sites. Besides, it is necessary to displace all atoms by ¼ along the *x* axis relatively to their values in the initial space group *Pbam*.

### 3.3. Magnetic ordering as the cause of structural instability of TbMn$_2$O$_5$ and BiMn$_2$O$_5$

We have shown above that for emerging of magnetic ordering in RMn$_2$O$_5$ it is sufficient to eliminate or transform the *J*1 and *J*2 couplings into the FM state. It is possible as a result of displacements of the O2 and O3 ions, which are located in critical positions and make substantial contributions to the AF-component of the above interactions, along the *a*-axis from the line -Mn1-Mn1-. However, the displacements of the O2 and O3 ions in this direction are not polar; they just allow eliminating dipole moments in the Mn$^{4+}$O$_6$ octahedra chain that results in fulfilling only one of the conditions for emerging of ferroelectricity in these compounds. Moreover, such displacements can be made in all three space groups under consideration, including the initial centrosymmetrical *Pbam* group.

Unfortunately, we do not still have reliable data on the changes in positions of not only oxygen anions, but also of manganese cations under effect of high magnetic fields. That is why to calculate the values of displacements of oxygen ions accompanying the emerging of the magnetically ordered state and ferroelectric polarization we used the elementary unit parameters and Mn ions coordinates obtained at temperature 27 K for TbMn$_2$O$_5$ in [3, 4] and at 100 K for BiMn$_2$O$_5$ in [11]. The initial oxygen cations for oxygen atoms were taken for TbMn$_2$O$_5$ from [35], for BiMn$_2$O$_5$ – from [11].

By varying the *x* coordinates of O2 and O3 ions, it is easy to calculate, by using the equation (2) and the "MagInter" software, their values at which the distances *h*(O2) and *h*(O3) would increase (by ~0.13-0.14 Å) up to 1.40 Å while the *j*(O2) and *j*(O3) contributions, respectively, would decrease down to 0 (see Section 3.1). According to the calculations, magnetic ordering is accompanied by increase of the *x* coordinate of the O2 ions by 0.019(0.018) and of the O3 ions – by 0.022(0.021) relatively to initial coordinates of these ions in the paraelectric phase of Tb(Bi)-compounds. Besides, in TbMn$_2$O$_5$ we displaced the O4 ion along the *c*-axis by 0.038 Å, thus increasing its *z* coordinate by just ~0.007, in order to decrease the probability of emerging competition of *J*1 and *J*2 with $J_3$ and $J_{3'}$ couplings. As a result of the performed displacements, the d(Mn1-O2) and d(Mn1-O3) bond lengths in the octahedron increased up to 2.013(2.047) Å and 1.968(1.968) Å, respectively, while the d(Mn2-O3) bond length in the pyramid, on the contrary, decreased down to 1.868(1.935) Å in Tb(Bi)-compounds. It resulted in partial equalization of the d(Mn1-O2) and d(Mn1-O3) lengths in Mn1O$_6$ octahedra. Another important effect of the displacements consists in equalization of the bond valences between the Mn1 and Mn2 atoms whose values became 3.42(3.29) and 3.56(3.32), respectively, while the initial values of bond valences for Mn1 and Mn2 equal to 4.01(3.86) and 3.18(3.11), respectively, were close to ideal values.

Therefore, emerging of magnetic ordering is accompanied by reduction of dipole moments of the Mn1 octahedra and charge disordering between the Mn1 and Mn2 positions.

Complete elimination of the Mn1O$_6$ octahedra dipoles accompanies the transition of the *J*2 couplings into the FM-state. The models of TbMn$_2$O$_5$ and BiMn2O5 compounds with the ordered magnetic structure and the absence of dipole moments in Mn1O$_6$ octahedra are marked as 'M-DO'.

For M-DO models of Tb- and Bi-compounds the structural parameters, bond distances and bond valences of Mn1 and Mn2 ($V_{Mn1}$, $V_{Mn2}$) are presented in Tables 1 and 2 while the parameters of magnetic couplings calculated in the centrosymmetrical space group *Pbam* – in Tables 3 and 4.

In such a transition the d(Mn1-O2) and d(Mn1-O3) bond lengths are equalized due to further displacement along the *x* axis (by ~0.009(0.016)) of only one O3 ion that increases (by 0.045(0.079) Å) the Mn1-O3 bond lengths up to 2.013(2.047) Å in Tb(Bi)-systems. As a result, the Mn1 charge decreases down to 3.30(3.07) Å while the Mn2 charge, on the contrary, increases up to 3.70(3.54) Å. In fact, there proceeds an exchange of bond valences values between the Mn1 and Mn2 positions, i.e. a new charge ordering emerges that is reverse to the initial one. As a result, the crystal structure becomes unstable, since for Jahn-Teller Mn$^{3+}$ ions the coordination surrounding in the form of flattened out octahedron is not advantageous [39] while for the Mn$^{4+}$ ions the coordination surrounding in the form of square pyramid is not characteristic.

According to the calculations, all the respective magnetic couplings at magnetic ordering accompanied by as decrease as complete elimination of octahedral dipole moments are virtually identical, except for the *J*2 couplings. In the first case the *J*2 couplings are eliminated while in the second case they are comparatively strong FM couplings. In the chain along the *c*-axis the nearest-neighbor *J*1 and *J*2 and the next-nearest-neighbor $J_2$, $J_3$, $J_{3'}$ and $J_4$ couplings do not compete with each other, since they become ferromagnetic. The parameters of strong AF intra-dimer *J*5 coupling, *J*3, *J*6 (along *b*-axis) and *J*4, *J*7 (along *a*-axis) couplings in the *ab* plane did not virtually change relatively to the parameters in the frustrated phase. Only two weaker AF couplings between the Mn2 ions – *J*8 (in the *ab*-plane) and $J_{c2}$ (located through the parameter *c*) – undergo the phase transition AF→FM with increasing interaction strength. As a result, in the magnetically ordered structure the competition in the *J*2*J*3*J*3, *J*2*J*4*J*4 *J*1*J*3*J*6, *J*1*J*4*J*7, *J*8*J*3*J*3, *J*8*J*6*J*6 and *J*8*J*10*J*11 triangles disappears; however, a weak competition is preserved in the *J*9*J*4*J*4, *J*9*J*7*J*7 and *J*5*J*10*J*11 triangles, where the couplings are not equal in strength, and in the *J*9*J*10*J*11 triangle, where all three interactions are weak (figure 3).

*3.4. Ferroelectric transition as the way for removing structural instability*

It is possible to return a stable state to the magnetically ordered structure of Tb(Bi)Mn$_2$O$_5$ only by approximation of the bond valences of Mn1 and Mn2 to the initial values of ~4 and ~3, respectively. In other words, it is necessary to make the charge exchange or redistribute charges between Mn1 and Mn2. Under the magnetic ordering conditions and elimination of the octahedral dipoles, when the O2, O3 and Mn1 ions are fixed, it is possible exclusively by displacements of the O4, O1 and Mn2 ions along the *b*-axis. Transition from the centrosymmetrical *Pbam* into the non-centrocymmetrical space group *Pb2$_1$m*, allows, simultaneously with increasing the bond valence of Mn1 and decreasing the bond valence of Mn2, emerging of spontaneous electrical polarization due to displacements of the Mn2, O1 and O4 ions along the b-axis, which are polar in this group (figure 2(c)). Transition into the space group *Pba2* does not produce the required result, since polar displacements in this group comprise those along the *c*-axis; however, they cannot effectively change the Mn1-O4 and Mn2-O4 bond lengths and, respectively, the bond valences of Mn1 and Mn2 and stabilize the structure.

In the displacement-type ferroelectrics polarization is, as a rule, related to the cation displacement from the center of its surrounding oxygen octahedron, while the positions of all other atoms remain unchangeable, and has the same direction as displacement. Transition of the non-polar modification of the M-DO model into the polar modification is complicated by two circumstances. First, the restructuring must be accompanied by specific changes in bond valences of Mn1 and Mn2 (see section *3.3.*). Second, the Mn2 cations in the non-polar modification of the M-DO model are already located not in the center of a square pyramid. That is why to separate the gravity centers of positive and negative charges in the M-DO-model structure the displacements of only cations are not sufficient.

We suggest two polar models P-Mn2 and P-O1 within the frames of the space group *Pb2$_1$m*. In both models the polarization effect was achieved by reducing the lengths of the Mn2-O1, Mn2'-O4' and Mn1-O4 bonds oriented along the *b*-axis and increasing the lengths of the Mn2'-O1, Mn2-O4 and Mn1-O4' bonds oriented along the same axis, but antiparallel to the shortened bonds (tables 1, 2; figure 2(c)). During the development of polar models the ions displacements were performed relatively to their positions in the M-DO-model.

The polar P-Mn2 model for Tb(Bi)-systems is formed by displacements of both Mn2 and Mn2' cations (by ~0.02(0.05) Å) at equal distances along the *b*-axis in the same direction and the O4 anions displacements (by ~0.13(0.18) Å) and O4' (by ~0.11(0.12)Å) along the *b*-axis as well, but in opposite directions and at different distances. As a result of the performed displacements, the bond valence of Mn1 increased by 0.45(0.54), and the bond valences of Mn2 and Mn2' decreased by 0.30(0.40) in Tb(Bi)-systems and noticeably approached the ideal values. The polar model P-O1 was developed only due to the oxygen anions displacements. We just substituted the Mn2 and Mn2' displacements in the P-Mn2 model by those of the O1 anions of the same value, but in the opposite direction, and decreased the O4' ions displacements (down to ~0.09(0.07)Å in Tb(Bi)-systems). This resulted in the values of the Mn1, Mn2 and Mn2' bond valences approximately the same as in the P-Mn2 model (tables 1, 2).

Nevertheless, in both P-Mn2 and P-O1 models we achieved the effect of cations displacement along the *b*-axis in one direction: the Mn1 ion approaches the O4 octahedron vertex, the Mn2 ion – the O1-O1 pyramid edge, and the Mn2' ion – the O4'-O4' pyramid edge (figure 2c). If one changes the direction of ions displacement along the *b*-axis to the opposite, polarization will be changed by 180°: the Mn1 ion approaches the O4' vertex, the Mn2 ion – the O4-O4 edge, and the Mn2' ion – the O1-O1 edge. One should mention that in these models we deliberately increased the O4 ion displacement value more than necessary, as compared to that of the O4' ion, in order to demonstrate the possibility of participation of both Mn1 and Mn2 ions in polarization.

A principal difference between these models consists in the fact that the displacement of the Mn2 and Mn2' ions in the same direction along the *b*-axis in the P-Mn2 model results, in addition, in opposite in value the change of the Mn2-O3 and Mn2-O3' distances. This very fact could be the cause of the disappearing symmetry center in the crystal structures of $TbMn_2O_5$ and $BiMn_2O_5$.

The electric polarization is induced and maintained by magnetic ordering which emerges and exists under external magnetic field. According to the calculations, the parameters of magnetic couplings in polar P-Mn2 and P-O1 models are virtually identical to respective values in the M-DO model with ordered magnetic structure (tables 3 and 4).

## 4. Conclusions

We have shown the role of crystal structure in emerging of magnetic ordering and electric polarization in $TbMn_2O_5$ and $BiMn_2O_5$. According to the calculations, the sign and strength of magnetic couplings found on the basis of structural data, the paraelectric phase of $TbMn_2O_5$ and $BiMn_2O_5$ is a frustrated antiferromagnetic. In linear chains of $Mn1^{4+}$ along the *c*-axis the nearest AF *J*1 and *J*2 couplings compete with the next-to-nearest-neighbor $J_n$ couplings. In addition, the *J*1 and *J*2 couplings compete with all strong AF *J*3, *J*4, *J*6 and *J*7 couplings between linear chains of $Mn1^{4+}$ and dimers of $Mn2^{3+}$. The elimination or transformation the *J*1 and *J*2 couplings into the FM-state is sufficient for emerging of the magnetic ordering in $Tb(Bi)Mn_2O_5$. The latter is possible through slight displacements of intermediate ions (O2 and O3 ions from the line -Mn1-Mn1- along the *a*-axis and O4 ions along the *c*-axis), which are in critical positions. However, these displacements accompanying magnetic ordering are not polar; depending on the value, they just induce equalizing (charge disordering) or values exchange (new charge ordering) of the bond valences between the Mn1 and Mn2 ions, thus creating instability of the crystal structure. To approximate again the bond valence of Mn1 and Mn2 to the initial value under the magnetic ordering conditions is possible only due to displacement of Mn2 (or O1) and O4 ions along the *b* axis that is the cause of ferroelectric transition.

To sum it up, the fundamental cause of multiferroicity in the compounds under study is magnetic ordering of $Tb(Bi)Mn_2O_5$ under effect of external magnetic field accompanied by charge disordering that, in turn, induces the structure instability. In the second stage, only the charge ordering transition takes place, thus inducing ferroelectricity and restoring the structure stable state while preserving magnetic ordering. Our studies are in agreement with the recent work by Brink and Khomskii [40], where the generic mechanisms by which charge ordering can induce ferroelectricity in magnetic systems are presented. To obtain direct experimental evidence of the presence of structural changes accompanying the spontaneous polarization, one should perform diffraction studies of induced multiferroics exclusively under high magnetic fields.